\documentclass[prd,twocolumn,showpacs,nofootinbib,preprintnumbers,amsmath,amssymb]{revtex4}

\usepackage{graphicx}
\usepackage{dcolumn}
\usepackage{bm}
\usepackage{feynmp}
\usepackage{amsmath}
\usepackage{amssymb}
\usepackage{url}

\usepackage{color}

\newcommand\rhocrit{\rho_{\mathrm{c}}}

\begin{document}

\title{Dark matter with two- and many-body decays and supernovae type Ia}

\author{Gordon Blackadder}
 \email{blackadder@brown.edu}

\author{Savvas M. Koushiappas}
\email{koushiappas@brown.edu}
\affiliation{Department of Physics, Brown University, 182 Hope Street, Providence, Rhode Isand, 02912, USA}

\date{\today}
             
\begin{abstract}
We present a decaying dark matter scenario where the daughter products are a single massless relativistic particle and a single, massive but possibly relativistic particle. We calculate the velocity distribution of the massive daughter particle and its associated equation of state and derive its dynamical evolution in an expanding Universe. In addition, we present a model of decaying dark matter where there are many massless relativistic daughter particles together with a massive particle at rest. We place constraints on these two models using supernovae type Ia observations. We find that for a daughter relativistic fraction of 1\% and higher, lifetimes of at least less than 10 Gyrs are excluded, with larger relativistic fractions constraining longer lifetimes.

\end{abstract}

\maketitle

\section{\label{sec:level1}Introduction}

Decaying dark matter has become a matter of considerable interest over the last few years.
It has been conjectured to answer specific questions related to cosmological large scale structure and unexpected observations of high-energy neutrinos, the positron fraction, and gamma-ray measurements.  Further motivation comes from particle physics models beyond the standard model as well as a basic desire to better understand the nature of dark matter.

Issues surrounding structure formation center on two problems. First is the so-called ``cuspy core'' issue where observation suggests galaxy cores have a constant dark matter density whereas simulations predict a cuspy density profile rising in the center \cite{Kaplinghat:2005aa, Gong:2008aa, Strigari:2007aa, DeLope-Amigo:2009aa, Peter:2010aa, Wang:2012aa, Wang:2013aa}. Second is the problem of missing satellites where dark matter-only numerical simulations predict a large number of dark matter halos present in the potential well of a host halo, while observations support the existence of roughly a factor of 5-10 fewer (e.g., in the Milky Way halo -- for a more in depth analysis of these two problems see \cite{Weinberg:2013aa}). Both of these problems have been shown to be potentially solved by N-body simulations of decaying dark matter  \cite{Wang:2014aa}.

From the observational point of particle astrophysics there are some rather interesting problems that may relate to decaying dark matter. For example, recently, IceCube reported on the observation of  very high-energy neutrinos ($\sim$PeV energies) whose observational properties (energies, directions, and flavors) are not consistent with what one would expect from the known backgrounds at the $4\sigma$ level \cite{IceCube-Collaboration:2013aa,2014PhLB..733..120E}. The now well-known problem of excess energetic positrons as reported by PAMELA \cite{Adriani:2011ab} and AMS-02 \cite{Aguilar:2013aa} may also have a decaying dark matter explanation (e.g., \cite{Ibarra:2014aa, Cirelli:2012aa}) as well as observations of gamma-ray lines and diffuse background measurements \cite{Yuksel:2008aa, Bell:2010aa,Buchmuller:2012aa}.
  
From the theoretical particle physics point of view, there are many dark matter candidates that arise in the context of decays in physics beyond the standard model, such as sterile neutrinos \cite{1994PhRvL..72...17D}, hidden photinos \cite{Morrissey:2009aa}, gravitino dark matter \cite{Moroi:1995aa} (all of which are discussed in detail in  \citet{Essig:2013aa}), as well as cryptons \cite{Ellis1990257}, moduli dark matter\cite{Asaka:1998aa}, axinos \cite{Kim200218}, and quintessinos \cite{Bi:2004aa} (all of which are  covered in \cite{Chen:2004aa}). Indeed as is pointed out by \citet{Ibarra:2013aa} there is no \emph{a  priori} reason to believe that dark matter particles should be absolutely stable.

One of the draw backs to considering specific dark matter candidates with particular decay channels and known outcomes is that while they can be tightly constrained such results are not widely applicable. Many  general decaying dark matter models have been derived that try to make only a few assumptions about decay products. 

In this paper we derive two rather general models. In the first we assume that the parent dark matter particle decays over time to a single, massless and relativistic particle and a single, massive and possibly relativistic particle (with velocity determined by momentum conservation). The velocity of the massive particle falls as the Universe expands and therefore there is a distribution of velocities as different heavy daughter particles will have been created at different times. The only assumption made in determining the evolution of the velocity distribution (besides the existence of such a two-body dark matter decay) is that the particles are noninteracting (i.e., no standard model interactions and no interactions among themselves). We also present a second model in which the assumption that there is only one massless particle is relaxed. This necessarily means that the velocity of the massive particle is indeterminate and so it will be assumed to be stationary. We then use recent  supernovae type Ia data to constrain these rather general models.

Type Ia supernovae are good candidates for constraining cosmological models. As standard candles their luminosity is well correlated with their observed brightness profiles. Therefore the only parameter affecting the observed luminosity flux is their luminosity distance -- the main idea behind the Hubble diagram and the revolutionary discovery of the accelerated Universe. Luminosity distance is a function of all known energy budget contributions to the Universe and their dynamics (including relativistic and nonrelativistic components). This fact is what motivates the use of supernovae type Ia as a probe of the possibility of a decaying dark matter scenario. Of course, as supernovae type Ia are late-universe standard candles [i.e., located at $z \sim {\cal{O}}(1)$], it is expected that their constraining power will be concentrated towards long-living decaying dark matter particles (of order the age of the Universe). 

The paper is structured as follows. Section \ref{sec:level2} derives the two-body decay while Sec.~\ref{sec:level3} looks at the many-body decay. Section \ref{sec:level4} details the data against which the models will be compared and considers some of the other relevant physics required to calculate the cosmological effects of decaying dark matter. The  results are given in Sec. \ref{sec:level5} alongside a detailed discussion on where these constraints fit in the bigger picture.

\section{\label{sec:level2}Two-Body Decay}

\begin{figure}[tbp]
\begin{center}
\includegraphics[width=0.25\textwidth]{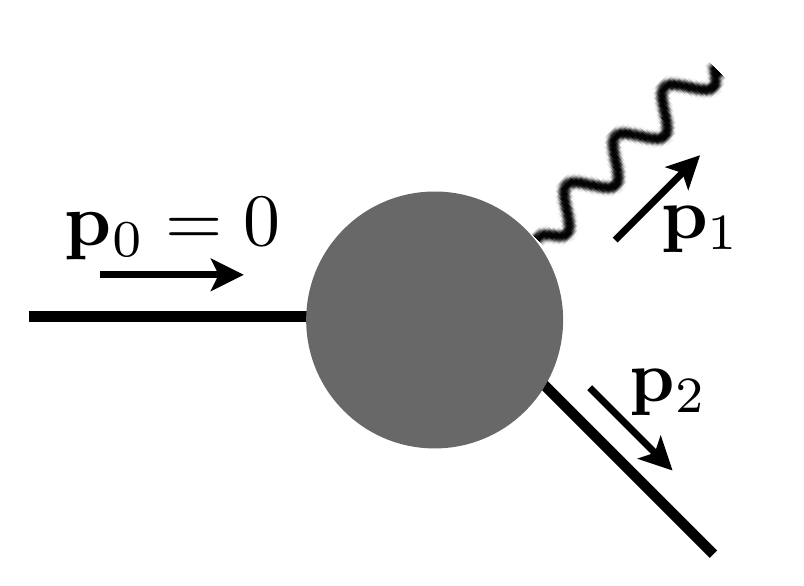} 
\caption{A pictorial of a two-body decay from a massive, stationary parent particle to a massless, relativistic particle and a massive, possibly relativistic particle.}
\label{fig:TWOpict}
\end{center}
\end{figure}

In this section we consider two-particle decay, with a parent dark matter particle (labeled with a subscript 0) with mass $m_{0}$ moving at rest relative to the expansion of the Universe, a massless and relativistic daughter particle (with subscript 1) and a second daughter particle (subscript 2) with mass $m_{2}$. The second particle may or may not be relativistic at the time  of its creation (see Fig.~\ref{fig:TWOpict}).

First, consider the 4-momenta of the particles at the time of decay, 
\begin{eqnarray*}
p_{\mu, 0} &=& (m_{0}c^{2}, \textbf{0}) \\
p_{\mu, 1} &=& (\epsilon m_{0}c^{2},\boldsymbol{p_{1}}) \\
p_{\mu, 2} &=& ((1-\epsilon) m_{0}c^{2},\boldsymbol{p_{2}}) 
\end{eqnarray*}
Here $\epsilon$ denotes the fraction of the energy of the  parent particle that has been transferred to the massless daughter particle. 
Energy and momentum conservation implies
\begin{equation}
\epsilon = \frac{\widetilde{m} \beta_{2}}{\sqrt{1-\beta_{2}^{2}}}, 
\label{momentum2body}
\end{equation}
and 
\begin{equation} 
(1-\epsilon)^{2}= \widetilde{m}^{2} + \frac{\widetilde{m}^{2}\beta_{2}^{2}}{1-\beta_{2}^{2}}
\label{energy2body}
\end{equation}
where $\widetilde{m} = m_{2}/m_{0}$ and $\beta_{2} = v_{2}/c$. Note that throughout the rest of the derivation we shall use natural units such that $c=1$. 

These two expressions give a unique relationship between $\epsilon$ and $\widetilde{m}$ and $\epsilon$ and $\beta_{2}$, 
\begin{equation}
\epsilon = \frac{1}{2}(1 -\widetilde{m}^{2})
\label{epsilon2body}
\end{equation}
\begin{equation}
\beta_{2}^{2} = \frac{\epsilon^{2}}{(1 - \epsilon)^{2}}
\label{beta2body}
\end{equation}
Note that $\epsilon = 0$ when $m_{2} = m_{0}$, and the maximum value is $\epsilon = 1/2$ when $\widetilde{m} = 0$. As $\epsilon$ approaches the value of 1/2, $\beta_{2}$ approaches the value of  1 which corresponds to  the second particle being relativistic with a boost given by
\begin{equation}
\gamma_{2} = \frac{1}{\sqrt{1 - \epsilon^{2} / (1-\epsilon)^{2}}}.
\end{equation}
We will now consider the evolution of the densities of these three particle species.

\subsection{The parent }

The rate of change of the parent particle is straightforward. The density decreases over time due to the expansion of the Universe and due to the decay. If the decay rate is $\Gamma = 1/\tau$ where $\tau$ is the lifetime of the particle, the time evolution of the parent particle is given by 
\begin{equation}
\frac{d\rho_{0}}{dt} + 3\frac{\dot{a}}{a}\rho_{0} = -\Gamma \rho_{0} 
\end{equation}
or
\begin{equation}
\rho_{0}(a) = {\cal{A}} \, \frac{e^{-\Gamma t(a)}}{ a^3},
\label{eq:parent2body}
\end{equation}
where ${\cal{A}}$ is some normalization constant and $t(a)$ is the age of the Universe at $a$. We choose to normalize the density of the heavy parent particle at the epoch of recombination (at a scale factor of $a_*$) using cosmic microwave background (CMB) data (see Sec. IV). 
We will further simplify matters by assuming that no decays occur in the early Universe before recombination. Under these assumptions the normalization constant is
\begin{equation} 
{\cal A} = \rhocrit  \Omega_{\mathrm{cdm}}  \, e^{ \Gamma t(a_*)}, 
\end{equation} 
where, $\rhocrit$ is the present value of the critical density, $\Omega_{\mathrm{cdm}}$ is the matter density as measured at the present epoch by CMB experiments, and $t(a_*)$ is the age of the Universe at the epoch of recombination (taken to be the age of the Universe that corresponds to approximately a redshift of $z \approx 1090$ \cite{Planck-Collaboration:2013ab}).

\subsection{The massless daughter}

The evolution of the massless daughter particle's density is governed by the decay rate of the parent particle and by the expansion of the Universe including the effect of redshifting, i.e., 
\begin{equation}
\frac{d\rho_{1}}{dt}  + 4 \frac{\dot{a}}{a} \rho_{1} =  \epsilon \Gamma \rho_{0}
\end{equation}
Using Eq.~(\ref{eq:parent2body}) we can write this as 
\begin{equation}
\rho_{1}a^{4} = \epsilon{\cal A}  \int_{a_*}^a \Gamma e^{-\Gamma t} a dt
\end{equation}
and with integration by parts and $d(e^{-\Gamma t}) =  -\Gamma e^{-\Gamma t} dt$ we get 
\begin{equation}
\rho_{1}(a) = \frac{\epsilon \cal A}{a^{4}} \left[ \int_{a_{*}}^{a} e^{-\Gamma t(a^{\prime})} da^{\prime} -  a^{\prime} \, e^{-\Gamma t(a^{\prime})} \bigg|_{a_{*}}^{a}\right], 
\end{equation}
where the lower bound of the integrals have been evaluated at $a_{*}$ in keeping with the boundary conditions. Evaluating the last term gives
\begin{equation}
\rho_{1}(a) = \frac{\epsilon \cal A}{a^{4}} \left[ \int_{a_{*}}^{a} e^{-\Gamma t(a^{\prime})} da^{\prime} -  a \, e^{-\Gamma t(a)} + a_{*} \, e^{-\Gamma t(a_{*})}\right]
\label{eq:rho1a}
\end{equation}

\subsection{The massive daughter and its equation of state}

Consider the change in the comoving abundance of the massive daughter particles at some time $t_{D}$ (or scale factor $a_D$). This is related to the comoving abundance of the parent particle by 
\begin{equation}
\frac{d  n_2}{dt_D} = - \frac{d n_0}{dt_D} = -\frac{d (a^{3}\rho_{0}/m_{0})}{dt_{D}} = \frac{\Gamma {\cal{A}} \, e^{-\Gamma t_{D}}}{m_{0}}.
\label{eq:n2numberdensity}
\end{equation}
In other words, the change in the number  of massive daughter particles is equal to minus the change in the number  of parent particles over  the same time interval; for every parent that decays one massive daughter is created.

The  momentum of the massive daughter particle at some later time $t>t_D$ when the scale factor is $a > a_D$ will then be inversely proportional to $a$ (i.e., the longer the particle has been around the slower it moves), 
\begin{equation}
p_2(a) =\frac{m_2 \beta_2 }{\sqrt{1 - \beta_2^2}} \, \left(  \frac{a_{D}}{a} \right).
\label{decayvelocity2particle}
\end{equation}
For small values of $\epsilon$, $\beta_2 \rightarrow 0$, and we recover the nonrelativistic redshifting of velocity ($v(a) \propto a^{-1}$).

The ratio of the energy of massive daughter particles at $a$ to the rest mass energy of the parent particle is 
\begin{eqnarray}
\frac{E_2(a, a_{D})}{m_{0}}&=& \widetilde{m} \sqrt{\frac{\beta_{2}^{2}}{1-\beta_{2}^{2}}\left(\frac{a_{D}}{a}\right)^{2}+ 1} \\
&=& \sqrt{1-2 \, \epsilon} \, \left[ \frac{\beta_2^2}{1 - \beta_2^2} \left(\frac{a_D}{a} \right)^2 + 1  \right]^{1/2}
\label{eq:particle2energy}
\end{eqnarray}

Equation~(\ref{eq:particle2energy}) shows that at early times ($a_D \approx a$) the energy of the daughter particle is as expected  $(1-\epsilon)m_{0}$, while at later times ($a_D \ll a$) it falls to $\sqrt{1-2\epsilon}m_{0}$. For small values of $\epsilon$ this effect is negligible, but as $\epsilon $ approaches the value of $\epsilon \approx 1/2$ this effect becomes significant as we will discuss further below. 

It is now relatively straightforward to derive  the energy density of massive daughter particles that were created at time $t_D$. First, we calculate the energy density at time $a > a_D$ as
\begin{eqnarray}
\rho_{2}(a) = \frac{1}{a^{3}} \int _{a_{*}}^{a} E_{2}(a, a_{D})dn_{2}(a_{D})
\end{eqnarray}
Substituting from Eq.~(\ref{eq:n2numberdensity}) and using $dt_D = da_D / (a_D H_D)$, where $H_D$ is the expansion parameter at the epoch of decay, we get that the total energy density of daughter particles at a redshift $a > a_D$ is
\begin{eqnarray} 
\rho_2(a) &=& \frac{ {\cal{A}} \, \Gamma \sqrt{1 - 2 \epsilon}}{a^3} \int_{a_*}^a  {\cal{J}}(a, a_{D})   da_D \label{eq:rho2evolution} \\
{\cal{J}}(a, a_{D}) & \equiv &  \frac{e^{-\Gamma t(a_D)}}{a_D H_D} \sqrt{ \frac{\beta_2^2}{1 - \beta_2^2} \left( \frac{a_D}{a} \right)^2 + 1 } \nonumber 
\end{eqnarray}
Note that the integral of Eq.~(\ref{eq:rho2evolution}) must be solved iteratively as $H_D$ [and consequently $t(a_{D})$] depends on the value of $\rho_2$ at each decaying epoch $a_D$ [$H_D^2(a_{D}) = H_0^2 \sum_i \Omega_i(a_{D})$, where $\Omega_i = \rho_i / \rhocrit$, and $i$ runs over all the constituents of the Universe, including the massive daughter particle (i=2) with density $\rho_2$]. 

Equation~(\ref{eq:rho2evolution}) shows that when $a\approx a_{D}$ and for large values of $\beta_{2}$ the density falls off as $a^{4}$ but as $a$ increases sufficiently ($ a \gg a_D$) the density falls with $a^{3}$. This is the case for a particle that is born relativistically at decay, but becomes nonrelativistic at late times (an important feature in decaying dark matter physics that has been relatively absent from the literature).

One convenient way of expressing the cosmological evolution of the massive daughter particle is the equation of state, 
\begin{equation} 
w_{2}(a) = \frac{1}{3} \langle v_2(a)^{2}\rangle. 
\end{equation} 
This useful quantity can be derived from some basic thermodynamic assumptions and the results of the previous section. 

\begin{figure}[]
\begin{center}
\includegraphics[scale=0.28]{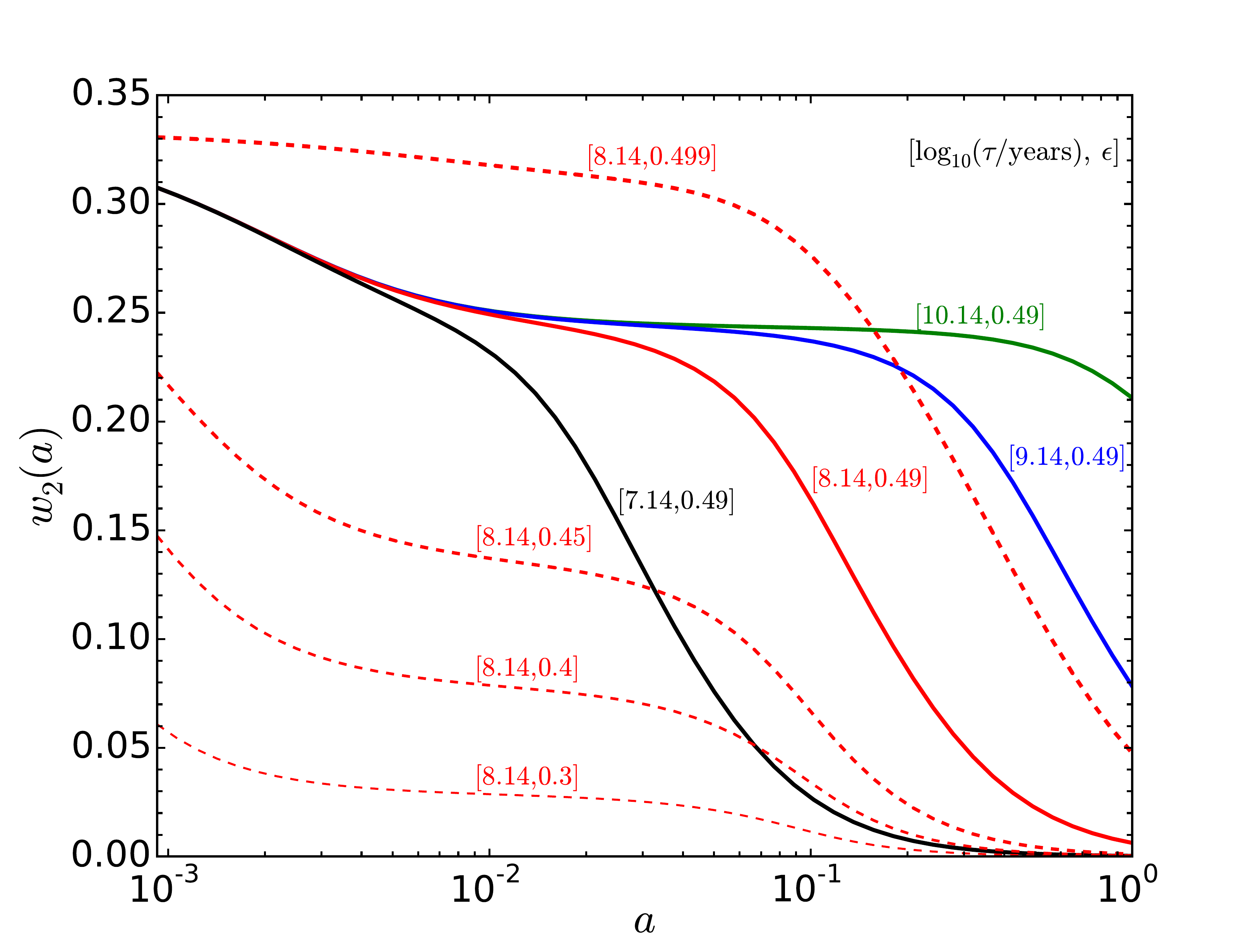}
\caption{The evolution of the equation of state of the massive daughter particle in two-body decays [Eq.~(\ref{eq:w2a})]. As the value of $\epsilon$ approaches $\epsilon \rightarrow 1/2$ the value of $w_{2}$ tends toward $w_2 \rightarrow 1/3$. Increasing the lifetime prolongs the period of time during which $w_{2}$ has an elevated value. Note that the current age of the Universe corresponds to the solid green curve with $\log_{\mathrm{10}}(\tau / {\mathrm{Gyr}}) = 10.14$.}
\label{fig:w2}
\end{center}
\end{figure}

The velocity of a massive daughter particle, whose parent decayed at $a_D$, has velocity $v$ at epoch $a$ given by
\begin{equation}
v_2^2(a,a_D) = \frac{ (a_D/a)^2 \beta_2^2 }{1 + \beta_2^2 [ ( a_D/a)^2 - 1]}
\end{equation}
The averaged (over all particles) velocity is derived by integrating over all particles that were created at or before $a$.
\begin{equation}
\langle v^{2}(a) \rangle = \left[ \int_{a_*}^a v^{2}(a,a_D) \frac{dn_{D}}{dt_D} \,   dt_D \right] \left[ \int_{a_*}^a \frac{dn_{D}}{dt_D}  \, dt_D \right]^{-1}
\end{equation}
We can use  Eq.~(\ref{eq:n2numberdensity}), and by expressing the integral over time as an integral over the scale factor in the first term we get 
\begin{eqnarray}
w_2(a) 	&=& \frac{1}{3}\frac{ \Gamma \beta_2^2}{e^{- \Gamma t_* } - e^{- \Gamma t} }\nonumber \\ 
		 & \times & \int_{a_*}^{a} \frac{ e^{-\Gamma t(a_D)} d\ln a_D}{H_D [(a/a_D)^2 ( 1 - \beta_2^2 ) + \beta_2^2]} \label{eq:w2a}
\end{eqnarray}
This expression must also be solved iteratively as $H_D$ is a function of $\rho_2(a)$ and its evolution with $a$ (thus $w_2(a)$). Figure~\ref{fig:w2} shows the evolution of the equation of state of the massive daughter as a function of scale factor for various values of $\epsilon$ and $\tau$. For the highly relativistic case where $\epsilon$ approaches the value of $1/2$ the massive daughter behaves at early times in a fashion similar to radiation (i.e., with a value of $w_2 \approx 1/3$). At later times (depending on the decay time scale) the value of $w_2$ decreases, thus the massive daughter behaves in a nonrelativistic manner, with an equation of state that approaches $w_2 \approx 0$ as expected. 

Note that Eq.~(\ref{eq:w2a}) and Fig.~\ref{fig:w2} serve as a sanity check on the validity of the calculation presented here; however, in practice it is easier to implement \ref{eq:rho2evolution} rather than \ref{eq:w2a} (as we discuss in the next section).

\section{\label{sec:level3}Many-Body Decay}

\begin{figure}[tbp]
\begin{center}
\includegraphics[width=0.25\textwidth]{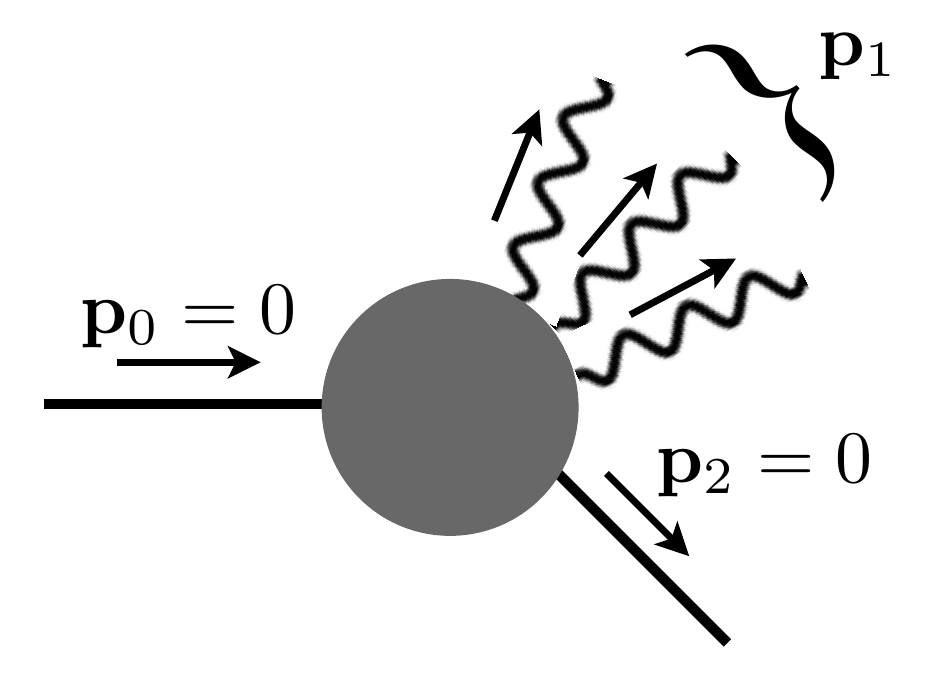} 
\caption{A pictorial of a many-body decay from a massive, stationary parent particle to many massless, relativistic particles and a massive, stationary daughter particle.}
\label{fig:MANYpict}
\end{center}
\end{figure}

In this section we discuss the relevant physics of many-body decay in which the daughter products consist of many relativistic particles and a single massive particle (see Fig.~\ref{fig:MANYpict}). By loosening the constraint on the number of relativistic particles we lose the ability to determine the velocity of the heavy daughter. For this reason we assume the particle to be stationary. Nonzero values of velocity could have been assumed (as e.g., in \cite{Bell:2010aa}) but this is necessarily arbitrary.

Parts of the derivation for the two-particle decay are identical to the many-body case. For simplicity the parent will still be referred to with subscript $0$, the massless daughter particles, even though there are many of them, with subscript $1$ and the heavy daughter with subscript $2$. Equation \ref{eq:parent2body} for the parent particle density is the same as is the formula for the relativistic particles in Eq.~(\ref{eq:rho1a}). Note however that this density refers to many relativistic particles created in each decay. 

Note also that in the many-body decay $\epsilon$ has a slightly different form. In the two-body decay $\epsilon$ was defined as the fraction of the energy of the parent particle that was transferred to the massless particle and  a formula was derived in terms of $m_{0}$ and $m_{2}$. We maintain this definition but in the case where there are many massless, relativistic particles and a single, stationary, massive particle formula is more simply derived as being
\begin{equation}
\epsilon = \frac{m_{0} - m_{2}}{m_{0}}
\end{equation}
where we see that $\epsilon$ is allowed to take any value between $0$ and $1$ (in contrast to the two-body case where the limit was $1/2$). 

The only particle density that has a different form in the many-body case is the heavy daughter which, without having to consider its kinetic energy in the derivation, is much more straightforward. The evolution of the density of the massive daughter is governed by 
\begin{equation}
\frac{d\rho_{2}}{dt} + 3 \frac{\dot{a}}{a} \rho_{2} = (1 - \epsilon)\Gamma\rho_{0}\\
\end{equation}
whose solution is 
\begin{equation} 
\rho_{2} =\frac{{\cal{A}} ( 1 - \epsilon)}{a^3}     \left[  e^{-\Gamma t_*} -e^{-\Gamma t}   \right]
\label{eq:manyrho2}
\end{equation} 
where it has been assumed that $\rho_2 = 0$ at $t = t_*$. The evolutions of the parent $\rho_0$ and the relativistic by-products $\rho_1$ are governed by the same expressions as in the two-body, namely, Eqs.~(\ref{eq:parent2body}) \& (\ref{eq:rho1a}) respectively.  

\section{\label{sec:level4}Decaying dark matter and Supernovae Type Ia}

In the previous two sections we derived the dynamical evolution of two decaying dark matter models. Here we will explore the constraints on these models that come from the cosmological information encoded in the observed brightness of supernovae type Ia (SNIa). 

We use SNIa from the Union2.1 catalog of 580 supernovae \cite{Suzuki:2012aa}. The recently published Joint Light-curve Analysis  (JLA) catalog (from the SNLS-SDSS collaborative effort) \cite{Betoule:2014aa} features a greater number of supernovae as well as an improved photometric calibration of two of the largest supernova surveys. 
However, as of the time of writing the JLA collaboration has not yet published a data release that will allow the straightforward  propagation of statistical and systematic uncertainties and for this reason the Union2.1 data is used in this paper (as was the case with the Planck 2013 data release \cite{Planck-Collaboration:2013ab}). The Union2.1 data set includes 580 supernovae up to a redshift of about $z \approx 1.4$ and excludes those with redshift below $z=0.015$ in order to minimize any error due to peculiar velocities. 

The physically important quantity in SNIa is the luminosity distance as a function of redshift of each supernova (essentially a Hubble diagram). The luminosity distance is related to the redshift, in a flat Universe where the scale factor relates to the redshift via $a= 1 / (1 + z)$, by 
\begin{equation}
d_{L} (z) = \frac{c(1+z)}{H_{0}} \int_{0}^{z}  {\cal{F}}^{-1/2} (z^\prime) \, \,dz^{\prime} 
\label{eq:lumdist}
\end{equation} 
where 
\begin{eqnarray} 
{\cal{F}}(z^\prime) &=& \Omega_0(z^\prime) + \Omega_1(z^\prime) + \Omega_2(z^\prime) \nonumber \\
&+&  \Omega_\nu(z^\prime)  + \Omega_\gamma(z^\prime)  +  \Omega_{b}(z^{\prime}) + \Omega_\Lambda , 
\label{eq:Ffortwo}
\end{eqnarray} 
For each species $i=\{\gamma, \nu, 0, 1, 2, b, \Lambda\}$, $\Omega_i = \rho_i / \rho_{\mathrm{crit}}$, and $\rho_{\mathrm{crit}}$  
is the critical density of the Universe. The values of $\rho_0$ and $\rho_1$ are given by Eqs.~(\ref{eq:parent2body}) \& (\ref{eq:rho1a}) respectively. The evolution of $\rho_2$ is given by Eq.~(\ref{eq:rho2evolution}) for the two-body scenario and by  Eq.~(\ref{eq:manyrho2}) respectively for the many-body decay.  

Note that the redshift dependence of each dimensionless cosmological parameter $\Omega_i$ above is due to the fact that the abundance of each species changes with time  due to decay (for $i=0, 1, 2$) as well as the expansion of the Universe. This is especially important for $i=2$ where the redshift evolution contains both the effects of production (by decay) and the dynamical evolution of the population, some of  which may or may not be relativistic. 

The distance modulus is simply a manipulation of the luminosity distance (where $d_{L}$ is measured in parsecs):
\begin{equation}
\mu(z) =  5\log_{10}d_{L}(z)-5, 
\label{modulus}
\end{equation}
which is the parameter that is constrained by observations. 

We obtain goodness of fit constraints to decaying dark matter models parametrized by $\epsilon$ and $\tau$ in the following way. We  compute the luminosity distance to the $j$th SNIa with redshift $z_j$, and the subsequent distance modulus $\mu(z_{j, {\mathrm{DDM}}})$. We then compare that with the observed absolute magnitude and redshift of each SNIa. The sum of the squares of their variance weighted difference is the $\chi^2$ distribution of that particular dark matter decaying scenario. 
\begin{equation}
 \chi^{2} = \sum_{j=1}^{580} \left\{ \frac{1}{\sigma_{j}^{2}} \left[ \mu(z_{j, {\mathrm{DDM}}}) - \mu(z_j) \right]^2 \right\}
 \end{equation}
where $\sigma_{j}$ is the uncertainty of the distance modulus measured for each supernova \cite{Suzuki:2012aa} .
This is then compared to a $\chi^2$ distribution with 578 degrees of freedom (580 SNIa minus 2 degrees of freedom, corresponding to $\epsilon$ and $\tau$) and assign a goodness of fit confidence. 

In order to properly compute the luminosity distance in a decaying dark matter scenario via Eq.~(\ref{eq:lumdist}), we need knowledge of the rest of the cosmological energy budget (in addition to matter and radiation derived from the parent dark matter decay, either in the two-body scenario or the many-body scenario). 

The photon density $\rho_\gamma$ is derived from the present photon temperature $T_{\gamma,0} = 2.7255$K \cite{2009ApJ...707..916F} using $\rho_{\gamma}(a) = 4\sigma (T_{\gamma,0}/a)^{4}/c$, where $\sigma$ is the Stefan-Boltzmann constant. This temperature is inflated by electron-positron annihilation, a heating that did not affect the neutrino temperature which leads to the well-known result for massless neutrinos, $T_{\nu}(a) = \left(4 / 11\right)^{1/3} T_{\gamma}(a)$ and an energy density given by $\rho_{\nu}(a)  = N_{\mathrm{eff}} \left( 7 / 8 \right)  \left(4/ 11 \right)^{4/3} \rho_{\gamma}(a)$, 
where the effective neutrino number density, $N_{\mathrm{eff}}$, takes the standard value $N_{\mathrm{eff}}=3.046$ \cite{Mangano:2002aa}.

However, this standard treatment of neutrinos is slightly inaccurate because they are both relativistic and massive and therefore we follow Sec. 3.3. in \citet{2011ApJS..192...18K}  that provides the following expression for the energy density of massive neutrinos:
\begin{equation}
\rho_{\nu}(a) = \frac{7}{8} \left(\frac{4}{11}\right)^{4/3}N_{\mathrm{eff}}\rho_{\gamma}(a)  f(y)
\end{equation}
where
\begin{equation}
f(y) \equiv \frac{120}{7\pi^{4}} \int_{0}^{\infty} dx \frac{x^{2} \sqrt{x^{2} + y^{2}}}{e^{x}+1}
\end{equation}
This form of the neutrino density takes into account the transition from relativistic to nonrelativistic expansion. 
A fitting formula gives the approximation 
\begin{equation}
f(y) \approx [1 + (Ay)^{p}]^{1/p},
\label{eq:neuApprox}
\end{equation}
where $A = 180 \, \zeta(3) / 7\pi^{4} \approx 0.3173$, $p=1.83$ and $\zeta$ is the Riemann zeta function, which is what we use for the remainder of this paper.

As the decaying dark matter formalism that we derived in Sec.~\ref{sec:level2} is normalized to the value of dark matter at the epoch of the CMB we choose to use cosmological parameters from CMB experiments. 
We use the cosmological parameters derived from the combination of Planck \cite{Planck-Collaboration:2013ab} and low-l WMAP \cite{Hinshaw:2013aa} likelihoods with the high-l Atakama Cosmology Telescope \cite{Das:2014aa} and South Pole Telescope \cite{Reichardt:2012aa} likelihoods (which were combined in \cite{Planck-Collaboration:2013ab} and called in short {\it Planck + WP + highL}). These are: $\Omega_{\mathrm{CDM}} h^2 = 0.12025$, $\Omega_{\mathrm b} h^2 = 0.022069$, $h = 0.6715$, $z_* = 1090.43$ and $w = -1 $.  This cosmological model is consistent with the Union2.1 supernovae sample that we use here (see Fig. 19 in \cite{Planck-Collaboration:2013ab}, and we use it as a benchmark over which we can test the SNIa constraints on the two-body and many-body decaying dark matter scenarios \footnote{We can readily provide results upon request for many of the cosmological models discussed in \cite{PlanckSuppl}.}).

\section{\label{sec:level5}Results and Discussion} 
\begin{figure}[tbp]
\includegraphics[scale=0.28]{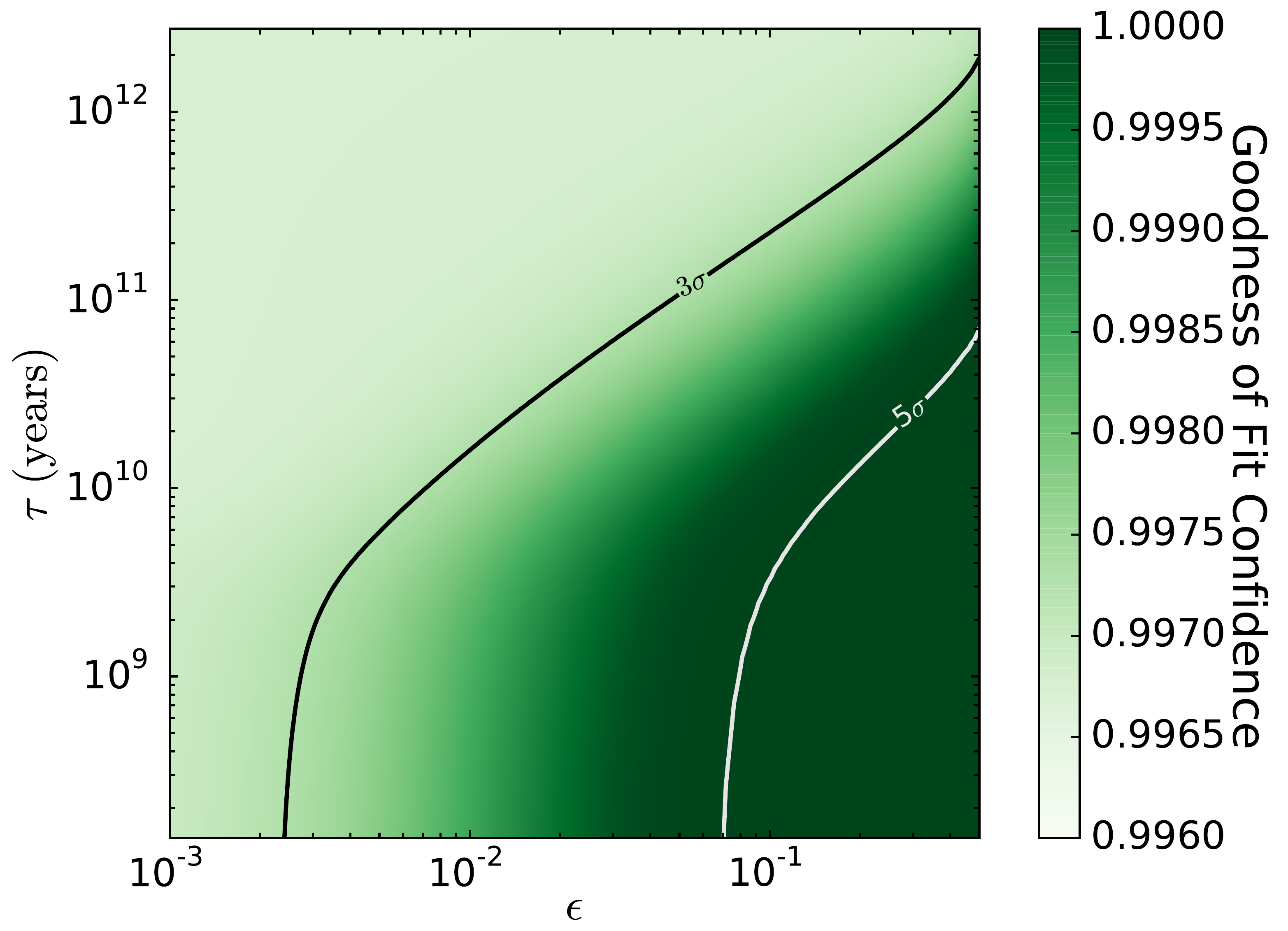} 
\caption{ Goodness of fit contour plots for the two-body decaying dark matter scenario in the $\epsilon -\tau$ parameter space. Color density corresponds to the value of the goodness of fit. The two contours depict the $3 \sigma$ and $5 \sigma$ values. The constraining power of supernovae is evident for lifetimes greater than $10^{10}$ years and values of the daughter relativistic fraction ($\epsilon$) greater than roughly $1\%$.}
\label{fig:tworesult}
\end{figure}

\begin{figure}[tbp]
\includegraphics[scale=0.28]{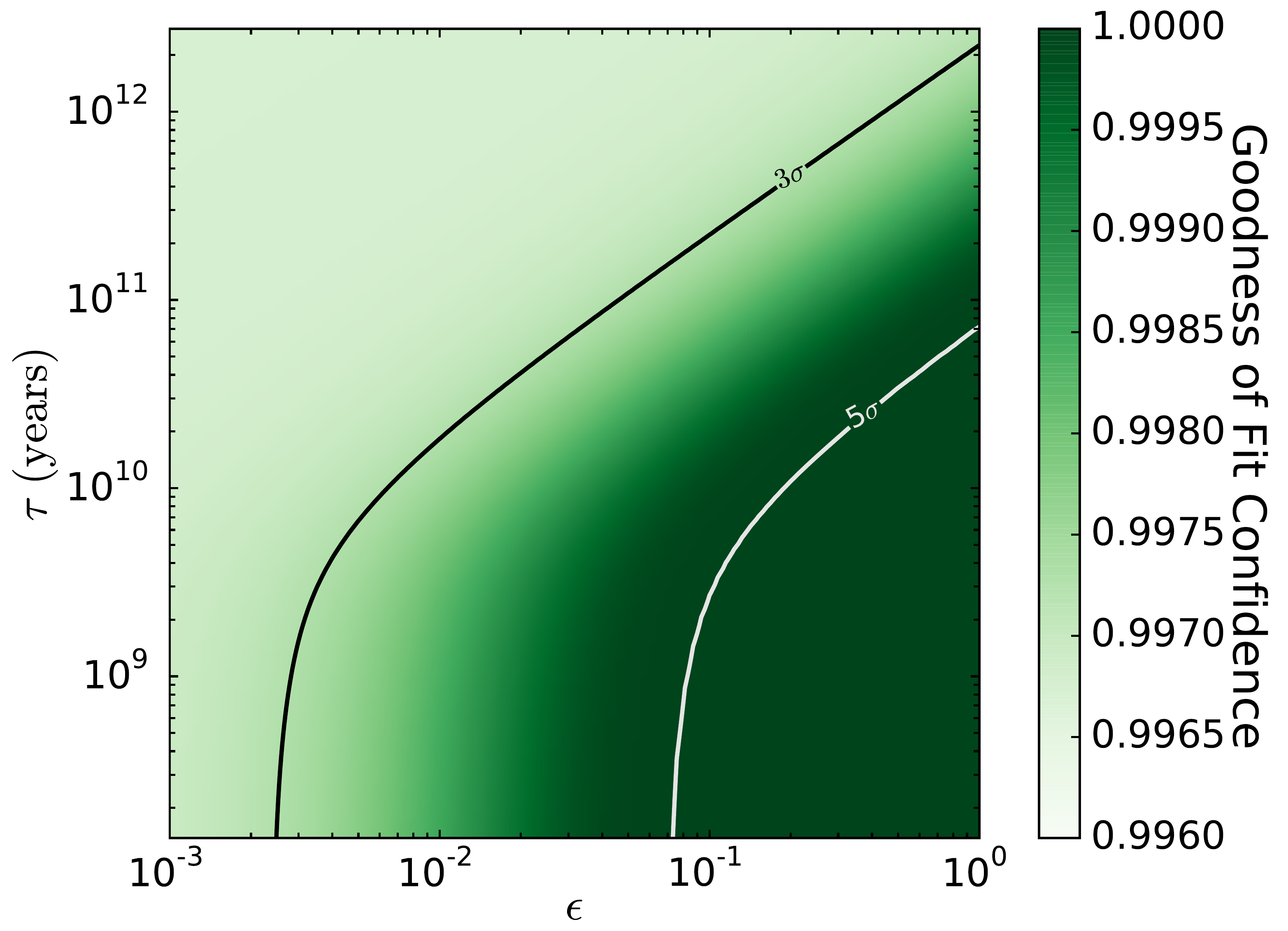} \\
\caption{ Goodness of fit contour plots for the many-body decaying dark matter scenario in the $\epsilon -\tau$ parameter space. Color density corresponds to the value of the goodness of fit. The two contours depict the $3 \sigma$ and $5 \sigma$ values. The constraining power of supernovae is evident for lifetimes greater than $10^{10}$ years and values of the daughter relativistic fraction ($\epsilon$) greater than roughly $1\%$.}
\label{fig:manyresult}
\end{figure}

Figures~\ref{fig:tworesult} and \ref{fig:manyresult} show the derived SNIa constraints on the two-body and many-body decaying dark matter scenarios, respectively. The color density corresponds to the value of the goodness of fit confidence, while the two curves depict the $3\sigma$ and $5\sigma$ contours in the $\epsilon - \tau$ parameter space. It is evident that the constraining power of SNIa is concentrated in large values of $\epsilon$ ($\epsilon > 10^{-2}$, and lifetimes of less than $\tau \sim 10^{10}$ years; the latter is not surprising as SNIa are fairly recent in cosmological history, and therefore only dark matter that decays appreciably at the sample epoch of the SNIa  we are using here can be constrained). 

Both plots look very similar and indeed share the same features (note that $\epsilon$ only extends to $1/2$ in the two-body case whereas it can rise as far as $1$ in the many-body scenario).
At short lifetimes, the contours are approximately vertical. The supernovae to which we are comparing only extend back to a redshift of $z \lesssim1.5$ and for very short lifetimes essentially all of the dark matter has decayed by this epoch rendering differences between small $\tau$ and even smaller $\tau$ irrelevant. 
Moving vertically up from small lifetimes to large lifetimes we see that the confidence level decreases. The longer the lifetime the smaller the difference between decaying dark matter and $\Lambda$CDM and so we essentially return to the base $\Lambda$CDM model found by Planck. Conversely moving horizontally from small $\epsilon$ to high $\epsilon$ the confidence level increases. As more and more radiation is added to the model the further away it is from the true Universe as traced by SNIa.  
At intermediate and high lifetimes we observe diagonal contours across the plots indicating that the effect of an increase in the lifetime (reducing the amount of additional radiation) can be offset by an increase in $\epsilon$. Finally observe that in the two-body case, but not the many-body, there is an upward inflection in the contour lines at high $\epsilon$. The reason for this can easily be seen by consulting Fig.~\ref{fig:w2} where $w_{2}$ varies greatly with changes in $\epsilon$ between $0.3$ and $0.5$.

It is important to also mention, however, that the choice of a cosmological model that sets the initial conditions can have a strong effect on the derived constraint, or turning the problem around, the results obtained here are rather sensitive on the choice of the cosmological model. For example, cosmological models that allow $w \neq -1$ have much more constraining power on decaying dark matter than models within the standard paradigm of~$w=-1$. In addition, if we use the WMAP9-only derived cosmological model \cite{Hinshaw:2013aa}, the constraining power of SNIa is less, scaling roughly by changing the $3\sigma$ contour into a $1\sigma$ contour. On the other hand, using the Planck-only cosmological parameters the constraining power of SNIa are stronger (perhaps a manifestation of the apparent tension between Planck and SNIa \cite{Planck-Collaboration:2013ab}). The choice of the aforementioned cosmological model of using Planck data together with  low-$\ell$ WMAP and high-$\ell$ ACT/SPT data ({\it Planck + WP + highL}) is however consistent with the Union2.1 supernovae we consider here and we feel this is the most appropriate and self-consistent choice of cosmological parameters in the normalization of the decaying dark matter models we explore here.

\begin{figure}
\includegraphics[scale=0.28]{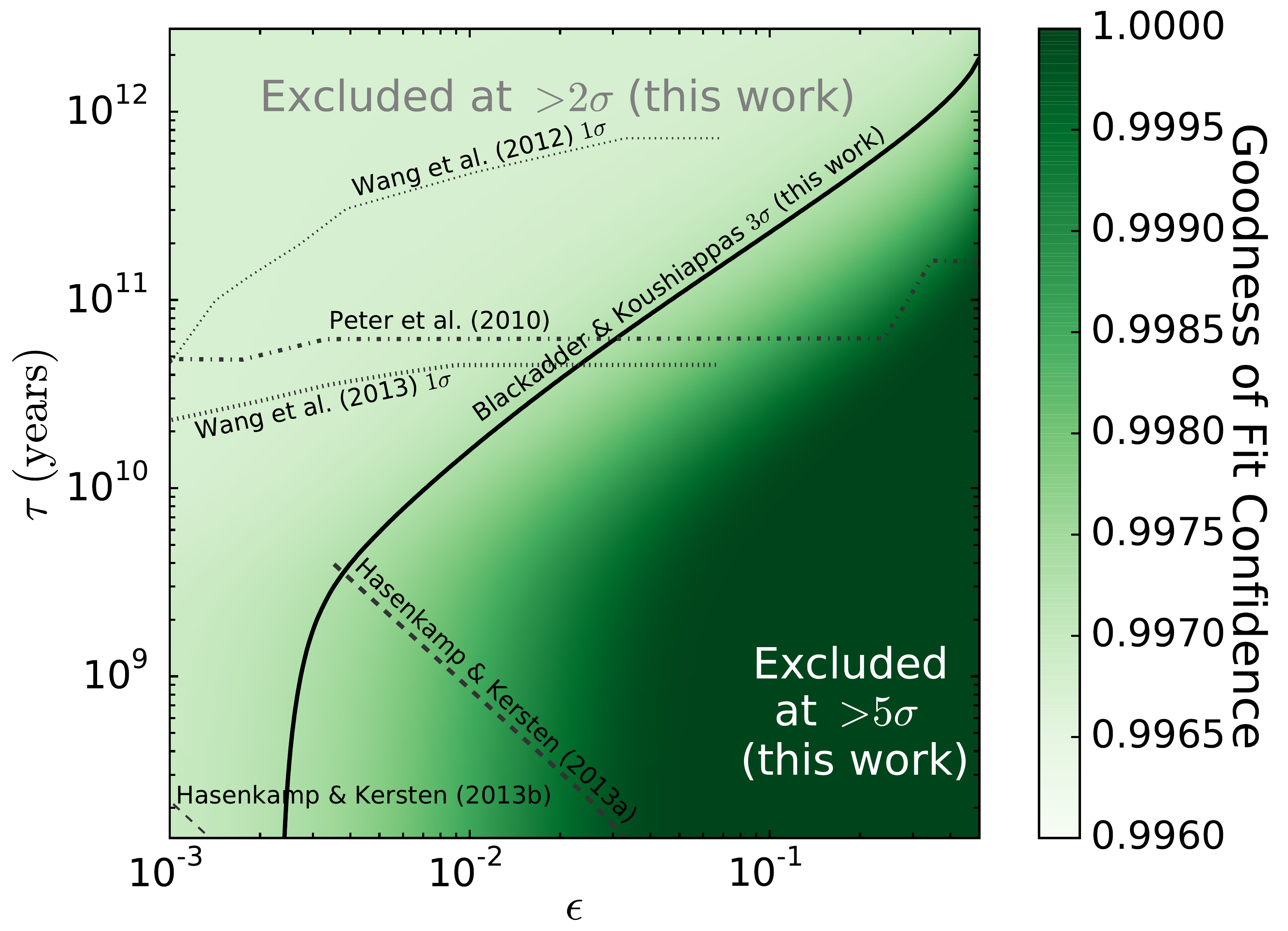} 
\caption{Summary of the two-body decay constraints presented here as compared to other studies. In all cases, parameter space is ruled out (at various levels of confidence) below the contour line. The results obtained in this work appear to rule out parameter space more aggressively than previous studies, but note that a direct comparison is not straightforward (for caveats see text). Note that both Wang et al. 2012 \cite{Wang:2012aa} and Wang et al. 2013 \cite{Wang:2013aa} considered only small mass splittings. This results in an abrupt cutoff in   their corresponding contour lines at lower values of $\epsilon$. Similarly  \citet{Hasenkamp:2013aa} considered only shorter lifetimes causing their contours to end abruptly in the parameter space.} 
\label{fig:twocompare}
\end{figure}

\begin{figure}
\includegraphics[scale=0.28]{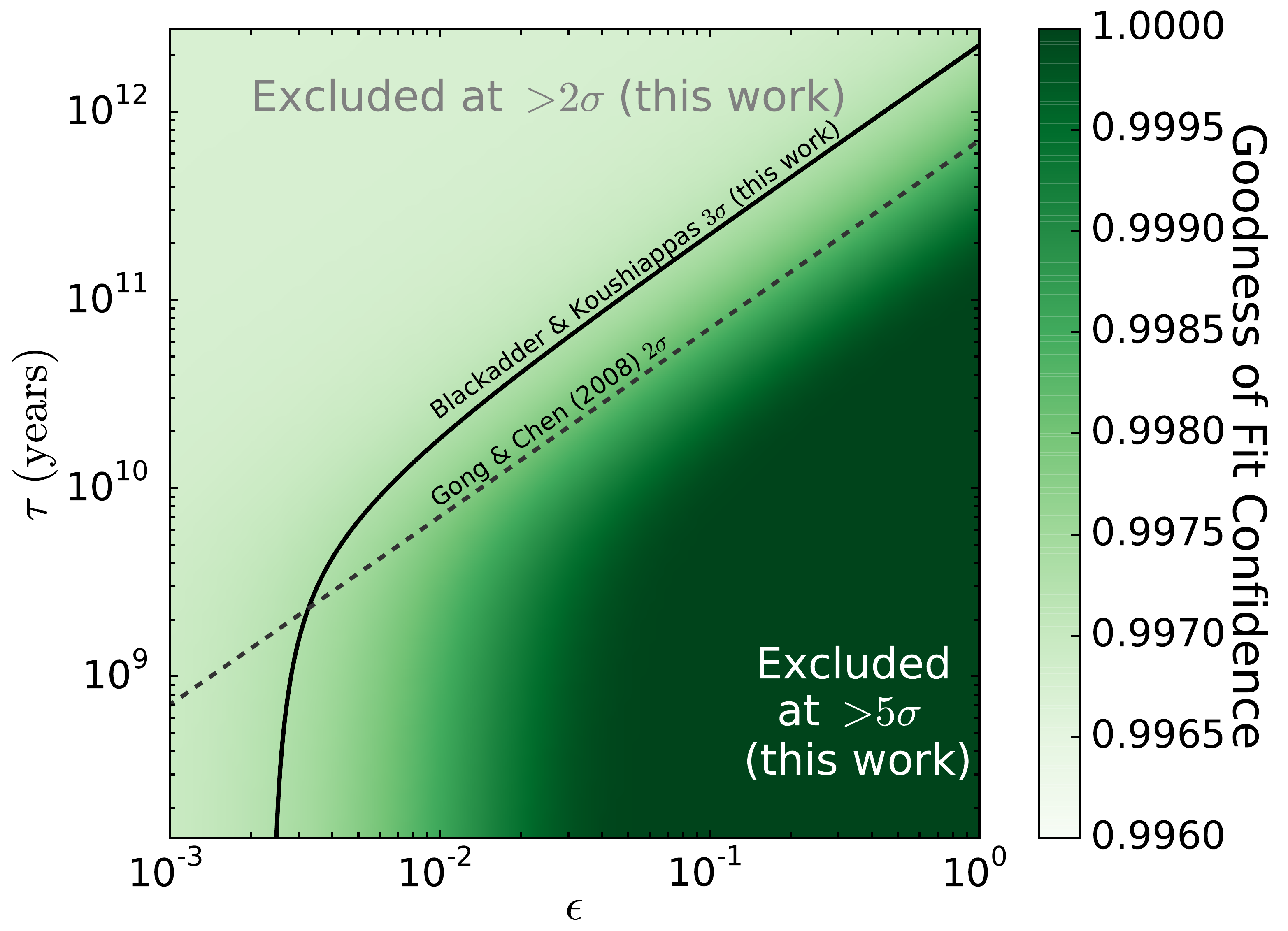} 
\caption{Summary of the many-body decay constraints presented here as compared to other studies. The results obtained in this work appear to rule out parameter space more aggressively than previous studies, but note that a direct comparison is not straightforward (for caveats see text).} 
\label{fig:manycompare}
\end{figure}

The derived constraints from SNIa on the two-body and many-body decaying dark matter scenarios are complementary to other approaches to the problem which we show in Figs.~(\ref{fig:twocompare}) and (\ref{fig:manycompare}). 

For example, in a recent paper, \citet{Hasenkamp:2013aa} derived a two-body decaying scenario with one daughter particle assumed to be of negligible mass and relativistic, and a second massive, possibly relativistic, daughter. They use a different, indeed complementary approach, to ruling out parameter space. They assume that the relativistic energy produced by decaying dark matter manifests itself as additional effective neutrinos, justified by findings such as in \citet{Dunkley:2011aa}. In addition, the  density of the decaying parent and daughters is allowed to vary between  models they explore, and is constrained by present limits on nonrelativistic and relativistic dark matter measurements. The density of the parent particle is allowed to vary between models in order to obtain the same amount of additional relativistic energy regardless of the other specified parameters, and they derive limits based on the current observed cold and hot dark matter densities. However, \citet{Hasenkamp:2013aa} make a number of simplifying assumptions. More specifically, they assume a sudden transition from radiation to matter domination, that the massive daughter particle is relativistic unless it's momentum is equal to or less than its mass, and that all the particles decay at a time equal to the lifetime $\tau$. This last assumption is obviously quite a simplification from the exponential decay and so the authors derive a correction factor to alter the density with two values, one when the lifetime is within radiation domination and one during matter domination. This approximation progressively improves for observational times significantly greater than the lifetime.   

Within this framework \citet{Hasenkamp:2013aa} looked at many different scenarios. For example they consider the contour where the number of additional neutrino degrees of freedom is 1 [labeled as Hasenkamp \& Kersten (2013a) in Fig.~\ref{fig:twocompare}].  They also considered a bound based on demanding that the amount of decaying dark matter could not exceed the total amount of dark matter that is observed [in their paper this was referred to as the non-domination constraint and in Figure~\ref{fig:twocompare} is labeled as Hasenkamp \& Kersten (2013b)].  Their results are complementary as they rule out parameter space at small values of $\epsilon$ and $\tau$ (the lower left area of the plot) while the results presented here, along with most previous findings, have ruled out space in the large values of $\epsilon$, small $\tau$ region (the lower right region). We note however that we do not make the assumption  of instantaneous decay and a relativistic cutoff of the heavy daughter as in \citet{Hasenkamp:2013aa}. Instead we allow both decay and relativistic behavior to be monotonically continuous functions, thus providing additional insight to the effects of decaying dark matter.

Another model of two-body decaying dark matter is considered in \citet{Yuksel:2008aa} in the specific case where the massless particle is a photon. However this model does not consider the (equal and opposite) momentum of the heavy daughter, merely constraining the decay by comparing the resulting photon density against the isotropic diffuse photon background and a Milky Way $\gamma$-ray line search. This model is further complicated as it only constrains the product $m_{\chi}\tau$, where $m_{\chi}$ is the mass of the parent particle, against the energy of the photon.  

Two-body decays in decaying dark matter were shown to be a possible solution to problems in structure formation in papers  by \citet{Kaplinghat:2005aa} and \citet{Strigari:2007aa} that looked at very early decays with lifetimes of less than one year and later decays ($z<1000$), respectively. They showed that dynamical dark matter could have positive implications for constant density cores in halos reducing the quantity of small scale substructure.

A pair of papers, \citet{Peter:2010aa} and \citet{Wang:2013aa}, analyzed structure formation data while looking at two-body decay scenarios where there was only a slight mass splitting between the parent and heavy daughter (a decay with small $\epsilon$) giving the massive particle a nonrelativistic velocity. The authors parametrized the decay in terms of the recoil ``kick'' velocity $v_{k}$ of the heavy daughter particle, which is given as $v_{k}/c \simeq (m_{0} - m_{2})/m_{0}$ where $m_{0}$ is the mass of the parent particle and $m_{2}$ is the mass of the heavy daughter. For the small values considered in these papers $\epsilon \simeq v_{k}/c$. Both of these constraints are shown in Fig.~\ref{fig:twocompare}. 

In \citet{Peter:2010aa} N-body simulations of dark matter halos are compared to observations of dwarf-galaxies, groups, and clusters to rule out regions of parameter space (note that this paper quotes a value of $\tau$ for different decay models which is the half-life of the decay as opposed to the lifetime used through out this paper). As this study was based on a suite of N-body simulations it is difficult to assign a numerical value of confidence. Instead, decaying dark matter parameters were allowed, if a few of the realizations of the satellite populations produced at least the minimum number of satellites expected in a Milky Way-like halo. 

Lyman-$\alpha$ forest data was used to constrain a dynamical dark matter model in \citet{Wang:2013aa}. Decaying dark matter affects structure growth and thus the authors used SDSS 1D Ly$\alpha$ data to measure large-scale structure growth \cite{McDonald:2006aa}.They looked at kick velocities up to $2\times 10^{7}$m/s though without considering relativistic effects. A related paper  by \citet{Wang:2012aa} projected how dynamical dark matter might be constrained by weak lensing results from future experiments such as Euclid \cite{Refregier:2010aa} and LSST \cite{LSST-Science-Collaboration:2009aa}. Recently \citet{Wang:2014aa} produced the most sophisticated N-body simulations of galaxy formation assuming dark matter decay. They showed that problems associated with large-scale structure formation such as the missing satellites problems are largely solved for particular values of the lifetime of the decaying dark matter particle and the recoil kick velocity of the daughter. 

\citet{Hasenkamp:2014aa} recently explored a two-body decay with lifetimes less than 1500 years, in particular decays occurring before and during big-bang nucleosynthesis. What they found is that there was no difference in the value of $N_{\mathrm{eff}}$ or $m^{\mathrm{eff}}_{\mathrm{hdm}}$ for the massive relativistic daughter compared to thermally produced $\nu_s$HDM but that the temperature at which such particles became nonrelativistic differed by a factor of $\sim 2$. Such a difference could have an observable impact on the CMB. This presents an interesting avenue of future work as there is a possibility of connecting the effects on the CMB to late Universe probes (longer lifetimes), such as the work presented here. 

When considering a many-body decay, \citet{Gong:2008aa} modified CosmoMC to include dynamical dark matter, and ruled out parameter space by making comparison to the distance modulus of 182 supernovae and the position of the first peak in the WMAP3 angular power spectrum (comparison shown in Fig.~\ref{fig:manycompare}). Since this paper was written there has been much improvement in the quantity and quality of the data, in particular with the 580 supernovae in the Union2.1 catalog \cite{Suzuki:2012aa} and in the Planck 2013 results \cite{Planck-Collaboration:2013aa}. 

Further constraints are placed on many-body decay by \citet{Zhang:2007aa} by assuming that some portion $f_{\chi}$ of the decay products are released as electromagnetically interacting particles and that some portion of that, $f$, is then deposited in baryonic gas thus affecting both reionization and recombination. Unfortunately this model can only constrain the product $ff_{\chi}$ against $\Gamma$ and it is difficult to map in the $\epsilon - \tau$ parameter space. 

A later paper by \citet{DeLope-Amigo:2009aa} aimed to update the results of  \citet{Gong:2008aa} and \citet{Zhang:2007aa}. However they only considered the specific case where all the energy from decay was transferred to relativistic energy, the $\epsilon = 1$ scenario. When this was true they found that in the case where the fraction $f$ of energy then deposited in baryonic gas was negligible then the Integrated Sachs-Wolf effect \cite{1967ApJ...147...73S} constrained the lifetime to be over $100$ Gyr at $2\sigma$ confidence. For non-negligible deposition they found $(f\Gamma)^{-1} \gtrsim 5.3 \times 10^{8}$Gyr. 

Specific decay models, where the decay is assumed to produce particular standard model particles, are more highly constrained than the general models above. \citet{Ibarra:2014aa} sets competitive limits on the lifetime of the parent particle by making comparison to the recent AMS-02 data release \cite{Aguilar:2013aa}. They assumed decay products such as $b\bar{b}$, $e^{+}e^{-}$, $\mu^{+}\mu^{-}$, $\tau^{+}\tau^{-}$ and $W^{+}W^{-}$ and set lower limits on the lifetime in the region of $10^{7} \sim 10^{11}$ Gyr. \citet{Essig:2013aa} looked at the same decays and found similar constraints when making comparison to recent gamma-ray and x-ray data from the Fermi Gamma-ray Space Telescope \cite{The-Fermi-LAT-Collaboration:2012aa}, INTEGRAL  \cite{Bouchet:2008aa}, EGRET \cite{Strong:2004aa}, and HEAO-1  \cite{Gruber:1999aa}. Similar constraints were also found by \citet{Cirelli:2012aa} using Fermi, H.E.S.S. \cite{2012ApJ...750..123A}, and PAMELA \cite{2009Natur.458..607A,2009PhRvL.102e1101A,2010PhRvL.105l1101A,2011PhRvL.106t1101A}.

It is worth noting that the recently measured high-energy neutrino detections at IceCube \cite{IceCube-Collaboration:2013aa} have been hypothesized as originating from decaying dark matter (see for example \citet{Ema:2013aa}) though more data will be needed before limits on such decaying models can be set. 

An interesting extension to specific decays was investigated in \citet{Bell:2010aa} where they considered a three-body decay in which the daughters consisted of two electrons, two photons or two neutrinos plus a heavy daughter that possessed a kick velocity. As there were three particles there was no derivable value of this velocity so they assumed it moved in the range of $[5-90]$ km/s in order to derive lower lifetime limits very approximately in the region of $10^{2}$ - $10^{8}$ Gyr. 

Of course there could also be decay into nonstandard model particles such as gravitinos, gauginos, or sneutrinos (see for example \citet{Ibarra:2013aa}).

As shown in Figs.~\ref{fig:twocompare} and \ref{fig:manycompare}, the results presented here rule out regions of contour space at clearly defined levels of confidence and do so at much higher levels than previously achieved. We underline that the comparison is somewhat opaque. In the two-body case the model developed here is more highly developed with its  sophisticated treatment of the possibly relativistic, heavy daughter particle. On the other hand our results rely only on comparisons to supernovae while the other plotted results were compared against other, and in many cases, several other data sets. It would therefore be of interest to implement the derived two-body and many-body decay scenarios to a multitude of cosmological probes \cite{BKinprep}, as well as generic particle physics models (e.g., dynamical dark matter \cite{2012PhRvD..85h3523D,2012PhRvD..85h3524D}). 

In summary, we developed a sophisticated model of two- and many-body dark matter decays. In the case of the former it takes into account the gradual slowing, from relativistic to nonrelativistic, of the heavy particle without having to choose an arbitrary cutoff for what counts as a relativistic velocity. The level of confidence at which areas of the decaying dark matter parameter space is strongly constrained by SNIa shows that cosmological probes may in fact strongly constrain decaying dark matter scenarios. 

\acknowledgements
We acknowledge useful conversations with Alex Geringer-Sameth, Jasper Hasenkamp, Deivid Ribeiro and Andrew Zentner. We thank the referees for the constructive feedback that helped improve the content of the paper. SMK is supported by DOE DE-SC0010010, NSF PHYS-1417505 and NASA NNX13AO94G. G.B. is partially supported by NSF PHYS-1417505. S.M.K. thanks the Aspen Center for Physics for hospitality where part of this work was completed.

\bibliography{decayingDM.bib}

\end{document}